\documentclass[journal=jpclcd,manuscript=letter]{achemso}

\usepackage{graphicx}

 \usepackage{csquotes}
 \usepackage{amssymb}
\usepackage{wrapfig}
\usepackage{float}

\newcommand{\onlinecite}[1]{\hspace{-1 ex} \nocite{#1}\citenum{#1}}
\newcommand{\ud}{\mathrm{d}}

\author{Milva Celli}
\affiliation{Consiglio Nazionale delle Ricerche, Istituto di Fisica Applicata ``Nello Carrara", \\ via Madonna del Piano 10, I-50019 Sesto Fiorentino, Italy}

\author{Lorenzo Ulivi}
\affiliation{Consiglio Nazionale delle Ricerche, Istituto di Fisica Applicata ``Nello Carrara", \\ via Madonna del Piano 10, I-50019 Sesto Fiorentino, Italy}

\author{Leonardo del Rosso}
\email{l.delrosso@ifac.cnr.it} 
\affiliation{Consiglio Nazionale delle Ricerche, Istituto di Fisica Applicata ``Nello Carrara", \\ via Madonna del Piano 10, I-50019 Sesto Fiorentino, Italy}

\title{Raman Investigation of the Ice Ic -- Ice Ih Transformation }

\keywords{Water ice, Ice XVII, Raman spectroscopy, Cubic ice}

\begin{document}


 \begin{center}
 {\bf \large \date{\today} }
 \end{center}

\begin{tocentry}
\resizebox{5cm}{!}
{
 \includegraphics[viewport= 4cm 0.1cm 23cm 21cm]
{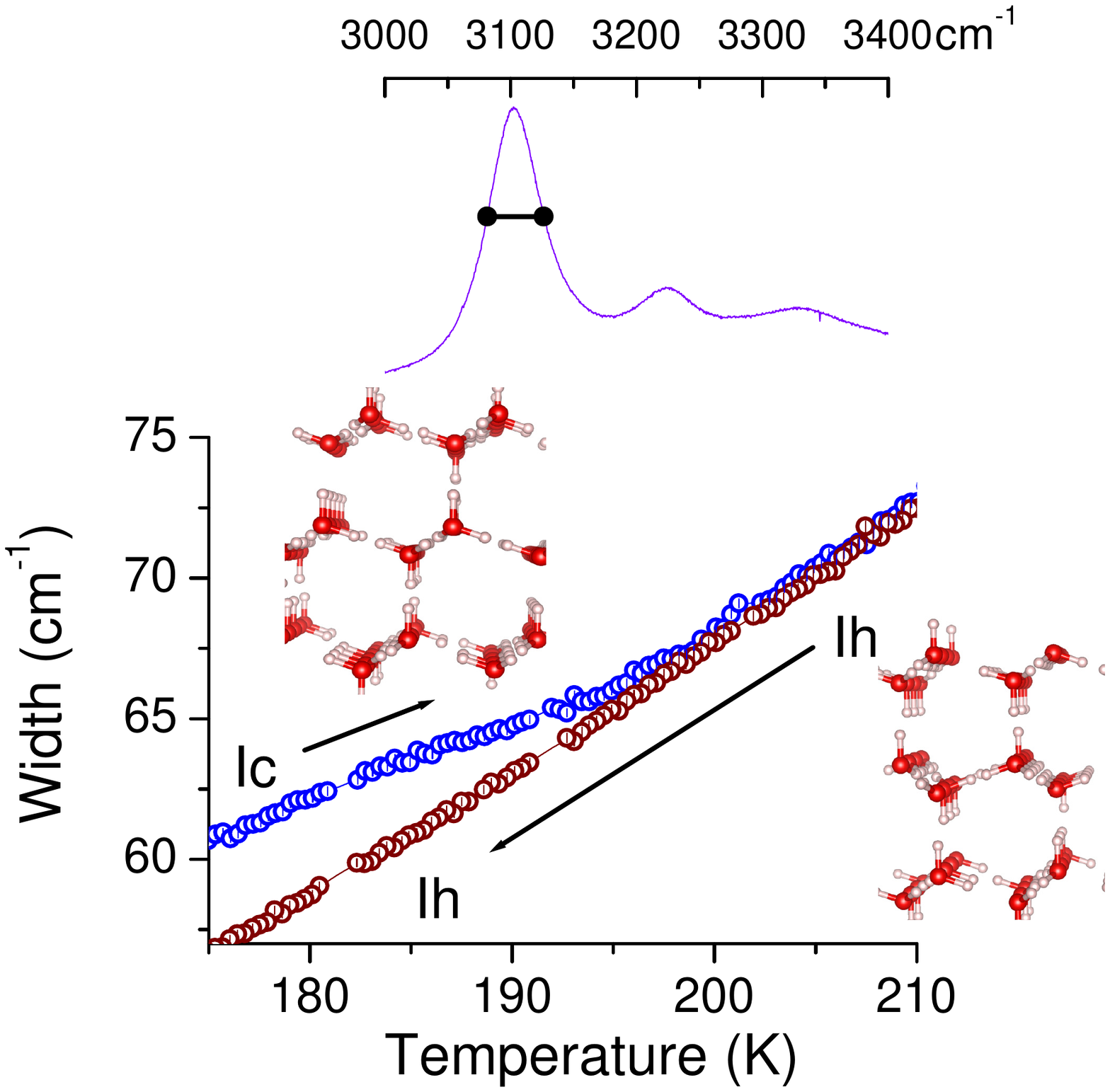}
}

\end{tocentry}

\begin{abstract} 

Among the many ice polymorphs, ice I, that is present in nature at ambient pressure, occurs with two different structures, i.e. the stable hexagonal (Ih) or the metastable cubic (Ic) one. An accurate analysis of cubic ice Ic was missing until the recent discovery of an easy route to obtain it in a structurally pure form. Here we report a Raman spectroscopy study of the transformation of metastable ice Ic into the stable form Ih, by applying a very slow temperature ramp to the ice Ic samples initially at 150 K. The thermal behavior of the spectroscopic features originating from the lattice and OH-stretching vibrational modes was carefully measured, identifying the thermodynamic conditions of the Ic--Ih transition and the dynamics of both structure with remarkable accuracy. Moreover, the comparison of two determinations of the transformation kinetics allowed also to provide an estimation of the activation energy for this transformation process.

\end{abstract}

The polymorphism of ice is an intriguing problem giving rise to one of the most vivid research topics in the physical-chemistry community to date. 
This is due not only to the key-role of water in every human activity, but also to the surprising complexity originating from a quite simple chemical system \cite{Petrenko99,Bartels-Rausch12,Salzmann11}.
The continuous research in this field has allowed to theoretically predict a plethora of stable and metastable crystalline ice phases, and experimentally determine nineteen different ice forms.
These have been found following different routes, e.g. applying extreme thermodynamic conditions \cite{Millot19}, doping ice forms with acids or bases to promote proton ordering \cite{Salzmann09} and evacuating gas hydrates \cite{Falenty14,del_Rosso16}.
At ambient pressure, only two forms of ice are thought to be present in nature on Earth: the ordinary hexagonal ice (ice Ih), and the cubic one (ice Ic).
While ice Ih is thermodynamically stable over a wide range of pressure and temperature and can be simply obtained by freezing liquid water,  cubic ice Ic can exist as a metastable crystal only at temperatures below about 200 K, but its importance for the physics of the atmosphere is relevant\cite{Whalley81,Murphy03,Murray15}.
 Ideally, the difference between these two structures can be understood considering the analogy with monoatomic systems, where a different sequence of hexagonally packed layers of atoms, stacked one on top of the other, can give rise to the cubic fcc or to the hexagonal hcp structure (with the ABCABC or ABABAB stacking sequence, respectively). In the case of ice, the stacking units are bi-layers of hydrogen-bonded water molecules, and the analogy with monoatomic systems is rigorous only if one neglects the hydrogen atoms.
 Since the middle of the last century,  numerous mechanisms were exploited to produce cubic ice,  as vapor deposition \cite{Konig43,Amaya17} or transformation of amorphous or crystalline high-pressure ice polymorphs \cite{Dowell60,Bertie63,Bertie64,Arnold68}.
However, all the samples of ``cubic ice'' obtained until recently are a stacking-disordered form of ice I (i.e. ice Isd), in which both hexagonal and cubic stacking sequences of hydrogen-bonded water molecules are present \cite{Kuhs12,Malkin12,Malkin15}, thus preventing an accurate measurement of the structural and dynamical properties of the pure cubic ice \cite{Salzmann19}.
Two recent and independent works have finally revealed the possibility to obtain cubic ice without stacking defects, either by heating over 150 K an ice XVII sample \cite{del_Rosso20} or by evacuating
 hydrogen gas from the H$_2$-H$_2$O hydrate initially in the C$_2$ phase \cite{Komatsu20}.
In particular, the method by del Rosso et al. \cite{del_Rosso20}, given the large availability of the starting material, i.e. ice XVII, represents the experimental keystone for different studies of ice Ic.

An important yet still unsolved issue is related to the difference of the dynamical properties of ice Ic and ice Ih, and whether spectroscopic techniques, in particular Raman scattering, can reveal signatures of either structures.
An attempt to identify differences in the O-H vibration Raman spectra between various instances of ice Isd and ice Ih has been reported by Carr et al. \cite{Carr14}. They have noted marked differences, positive or negative and in some cases up to 7 cm$^{-1}$, in the peak position of the OH stretching band measured at 80 K.
One could expect that similar differences might be present also comparing ice Ih with ice Ic, but this is not the case, according to the latest measurements performed on ice Ic\cite{del_Rosso20}.

In this paper we report the results of a spectroscopic analysis of the transformation of ice Ic into ice Ih, performed by the measurement of a long sequence of Raman spectra during a slow thermal treatment of the sample, initially in the ice Ic metastable phase.
Ice XVII (H$_2$O) was obtained from the deuterium-filled water hydrate, initially in the C$_0$ phase, by means of a thermal procedure similar to that reported in Ref. \onlinecite{del_Rosso16jpcc},  and described in more detail in the Supporting Information (SI).
The sample sits on the bottom of a vacuum--tight optical cell, provided with one glass window, under vacuum or in a controlled gas atmosphere, in thermal contact with the cold finger of a closed circuit refrigerator  able to reach 10 K.
A computer--driven thermal controller can regulate the temperature with a sensitivity of $\pm 0.01$ K and an accuracy of $\pm 0.1$ K\cite{Giannasi08}. 
We used an Ar ion laser at $\lambda = 514.5$ nm for the excitation of the Raman spectra, and a  1401 SPEX spectrometer equipped with a cooled Andor CCD detector mounted after the first spectrometer stage. 
To highlight the smallest changes in the spectral shape and position, possibly occurring during a thermal treatment of the sample, 10 minute spectra have been recorded in a continuous sequence, maintaining the spectroscopic apparatus tuned on one of the two chosen frequency regions, corresponding either to the lattice modes (120-350 cm$^{-1}$) or the OH streaching band (3000-3400 cm$^{-1}$).
The frequency axis was calibrated at the beginning and at the end of each measurement cycle, by means of a Ne spectral lamp with a reproducibility of the order of 0.2 cm$^{-1}$ and an absolute accuracy of 1.0 cm$^{-1}$ (see SI).
We have therefore used two different samples from the same batch for this experiment.
In both cases, the first thermal treatment  consisted of a heating ramp at a rate of $\Delta T/ \Delta t \simeq 0.037$ K/min from 150 to 250 K.
During this first heating we have observed clear spectroscopic evidences of the transition from ice Ic to ice Ih.
Subsequently, spectra are measured cooling again the same sample, transformed into ice Ih,  either at some selected temperatures between 160 and 210 K, in the case of the lattice band, or performing a cooling ramp, at the same rate, down to again $T \simeq 150$ K, for OH stretching spectra.

\begin{figure}[h!]
\begin{center}
 \resizebox{0.5\textwidth}{!}
{
 \includegraphics[bb=0.1cm 0.1cm 21cm 29.7cm, viewport= 1cm 1.5cm 20cm 29cm]  {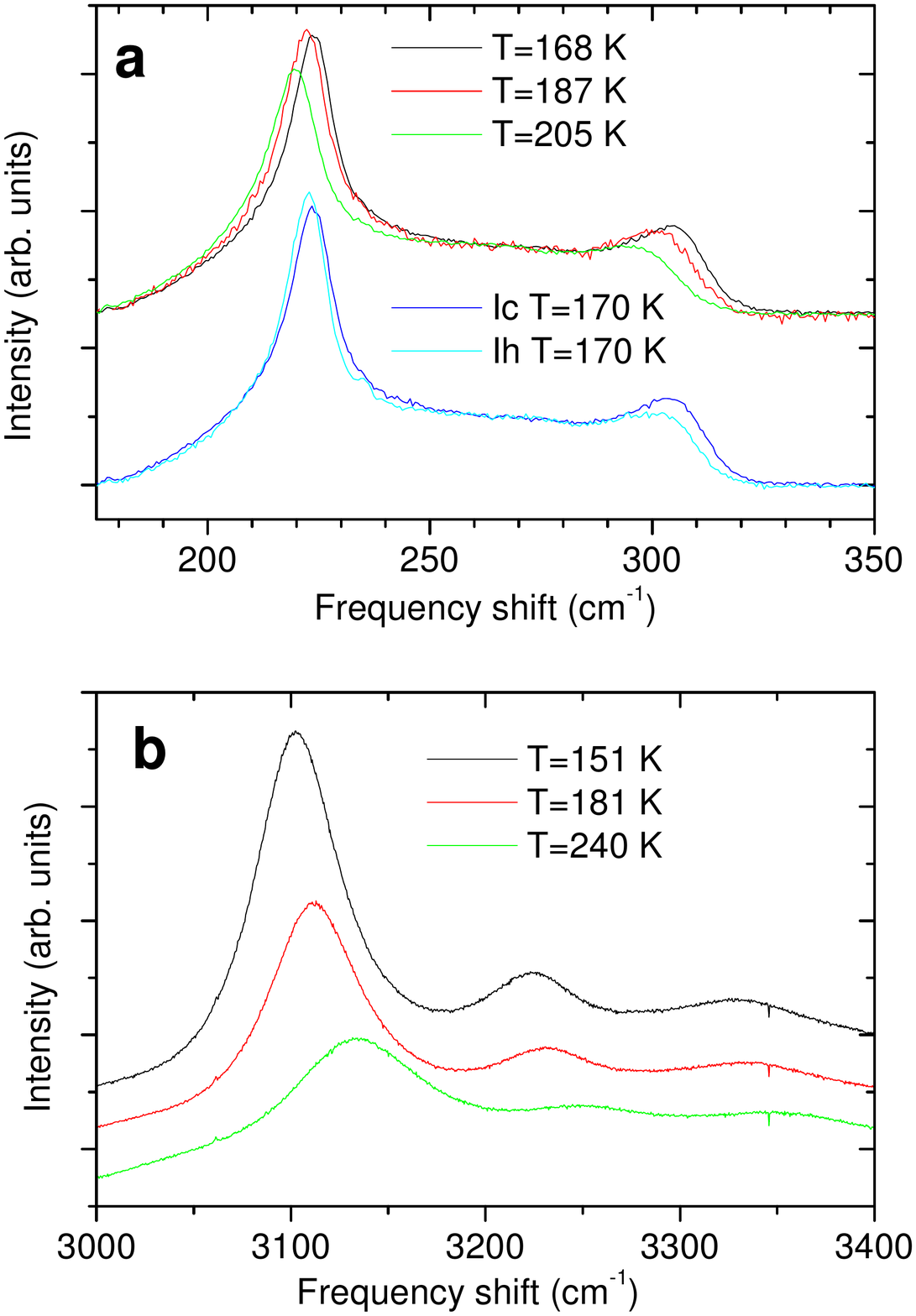}
}
%

\end{center}
 \caption{(a) Raman spectra in the lattice mode region, at three different temperatures, during the heating run (upper curves). 
 The blue and cyan spectra (lower curves) show the comparison of the spectra of ice Ic  and ice Ih,  at the same temperature $T= 170$ K.
 (b) Examples of spectra measured during heating in the OH stretching band region.
All spectra have been shifted for clarity.
}
\label{f.Spectra}
\end{figure}

\begin{figure}[h!]
           \begin{center}
 \resizebox{0.6\textwidth}{!}
{ \includegraphics[bb=0.1cm 0.1cm 21cm 29.7cm,viewport=1cm 2cm 19cm 29cm]
 {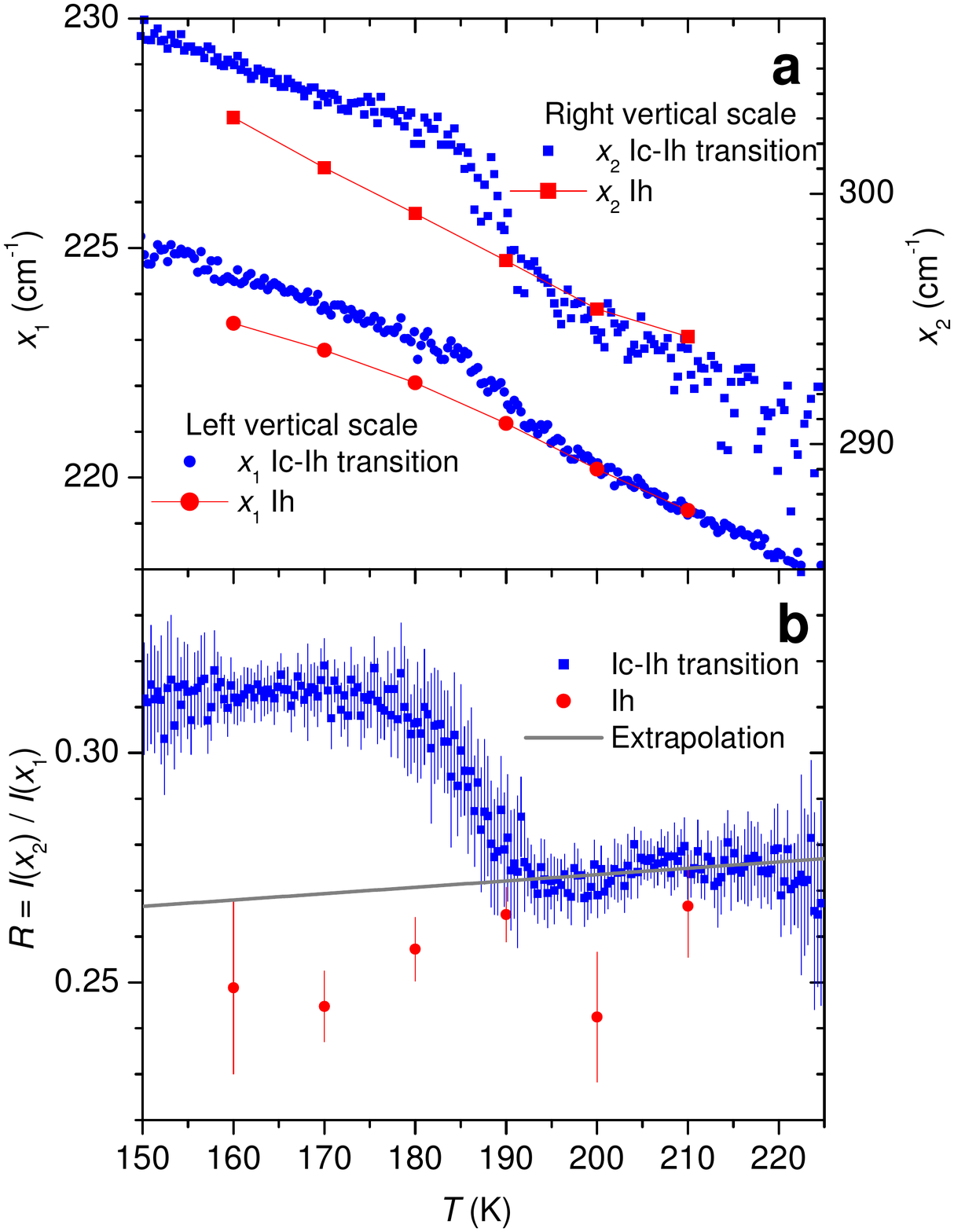}
}
\end{center}
 \caption{
  (a) Frequency positions $x_1$ and $x_2$  of the low-frequency and high-frequency maxima of the lattice mode band plotted as a function of temperature.
  (b) Ratio $R=I(x_2)/I(x_1)$ between  the spectral intensity at two maxima plotted as a function of the temperature. Blue symbols: measurements during the temperature ramp from 150 K to 250 K;  red symbols: measurements performed on the same sample after transformation into ice Ih. The solid gray line is a linear extrapolation of the data towards low temperature (see text).
}
\label{f.Lattice_XeR}
\end{figure}

Our preliminary Raman measurements on pure ice Ic\cite{del_Rosso20} indicated that notable differences with ice Ih are observed only at low temperature, and mainly in the lattice bands, while for the OH stretching mode only the width of the band is modestly affected.
Using the procedure described  in this work we have recorded  the transformation ice Ic - ice Ih while it is happening and measured its  kinetics.
The spectra (about 500 for each frequency region) have been analyzed by a computer routine, performing a fit of the spectral intensity with the aim of deriving characteristic parameters and measure their behavior with temperature.
Examples of the recorded spectra are shown in Fig.~\ref{f.Spectra}, for the lattice modes (a)  and the OH stretching band (b), where they have been shifted vertically for clarity.
The spectra in Fig.~\ref{f.Spectra}(a) have been normalized to have the same intensity at about 270 cm$^{-1}$, after the subtraction of a linear background, taken almost coincident with the raw signal at 175 and 325 cm$^{-1}$.
The main features observed are an evident peak at about 220 cm$^{-1}$ and a second maximum around 300 cm$^{-1}$.
The shape of the whole band changes when passing from ice Ic to ice Ih (Fig.~\ref{f.Spectra}(a), lower curves).
To quantify these changes as a function of temperature, we have recorded the frequency values of the two maxima, $x_1$ and $x_2$, and the ratio of the spectral intensity at these two frequency values, namely $R=I(x_2)/I(x_1)$.
These quantities are plotted, respectively, in Fig.~\ref{f.Lattice_XeR}(a) and Fig.~\ref{f.Lattice_XeR}(b), as a function of temperature, both during the heating ramp (blue symbols) and for spectra measured subsequently on ice Ih during cooling (red symbols).
From the observation of these figures, one can immediately conclude that the transformation from ice Ic into ice Ih, when applying such an heating rate, occurs at about 185-190 K. 

%

%
\begin{figure}[h!]
           \begin{center}
 \resizebox{0.6\textwidth}{!}
{ \includegraphics[bb=0.1cm 0.1cm 21.0cm 27.9cm,viewport= 1cm 1cm 20cm 27cm]{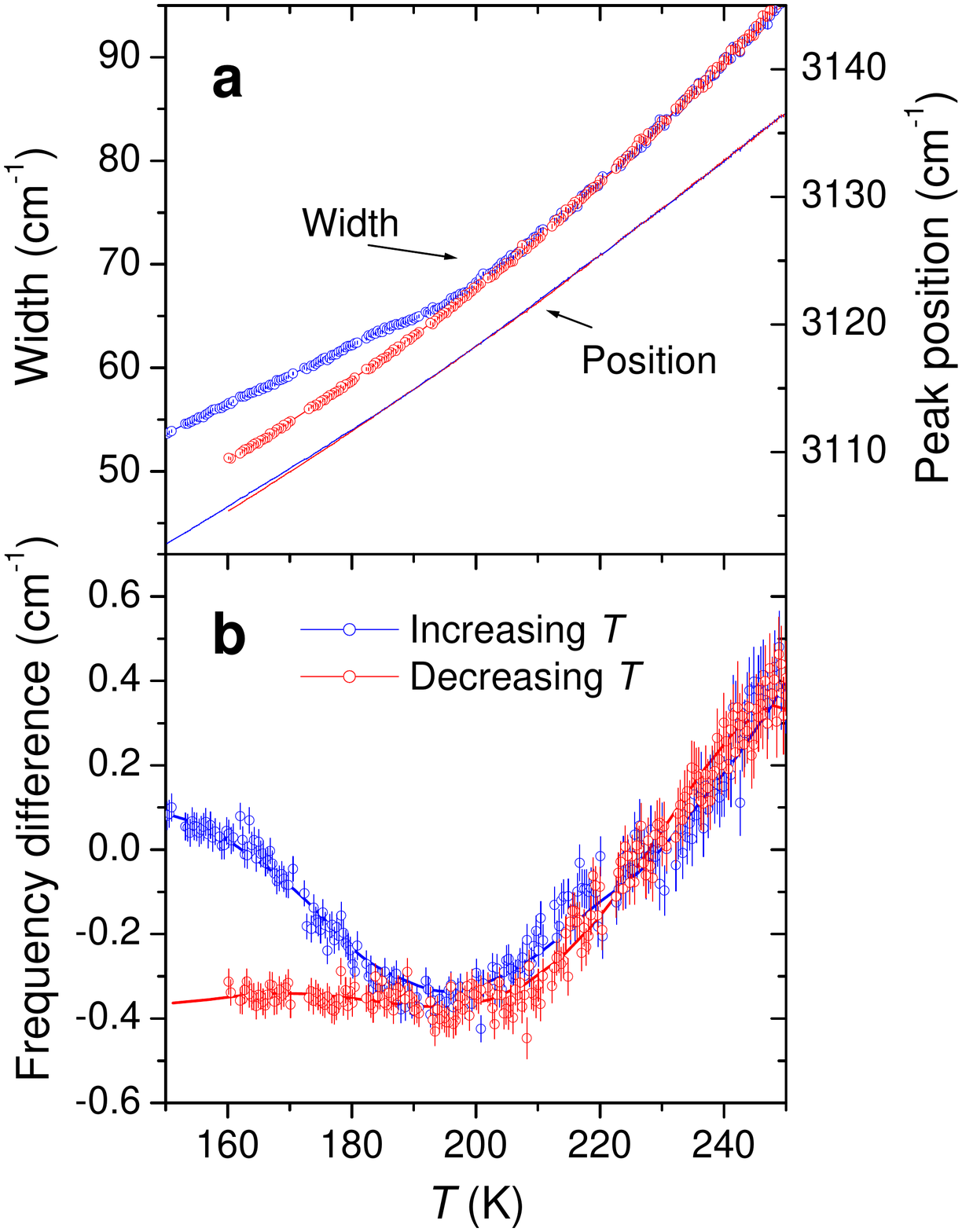}}
\end{center}
 \caption{(a) Temperature behavior of the position (right scale) and width (left scale) of the Lorentzian function modeling the lowest frequency peak of the OH stretching band, measured during heating (blue) and cooling (red) ramp.
The tiny difference in peak position visible for $T \lesssim 190$ K  is shown with more evidence in panel (b), where the peak frequency of the same band is reported, after subtraction of a smooth polynomial (see SI for details). 
}
\label{f.OH_param}
\end{figure}

%
The same transformation can be observed taking into account the spectra in the OH stretching region, which appear as formed by three bands (see Fig.~\ref{f.Spectra}(b)).
These has been fitted  with three Lorentzian over a linear background. Examples of the fit at the lowest and highest temperature are shown in Fig. 1 of the SI, where one observes a good fit in both cases. %
We have considered, for our analysis, the peak position and the width of the most intense band, the one at lowest frequency.
It is known that this peak moves sensibly towards high frequency with increasing volume\cite{Pimentel56,Pruzan94,Vos96}, as is reported in Fig.~\ref{f.OH_param}(a).
The  differences between the spectra measured during the heating and cooling ramp are very small, and are highlighted in Fig.~\ref{f.OH_param}, where we report, in panel (a)  the position and width (FWHM) of the Lorentzian line that fits the first peak and, in panel (b), the peak position after the subtraction of a smooth polynomial (see SI for details). 
The difference between ice Ic and ice Ih, magnified in Fig.~\ref{f.OH_param}(b), is quite small (0.4 cm$^{-1}$ maximum), but still detectable outside statistical errors, since the  reproducibility of our measurement is better than 0.2 cm$^{-1}$.
From these graphs one can measure a transformation temperature of about 190 K, in accordance with the results relative to the lattice mode.

The numerous Raman OH vibration spectra measured during this study involved several different spectra of ice Ih,  at temperatures between 10 and 250 K.
All these spectra has been analyzed by means of the fitting procedure described previously, and have allowed to determine, with an overall accuracy of $\pm 1$  cm$^{-1}$, the temperature shift of the peak position of the OH band, in a large temperature interval.
In the SI we present a functional form fitting  our measurements, discuss its  accuracy and compare it with several other determination in the literature.

\begin{figure}[h]
\begin{center}
 \resizebox{0.6\textwidth}{!}
 {
 \includegraphics[bb=0.1cm 0.1cm 21.0cm 27.9cm, viewport= 1cm 1cm 21cm 27cm]
 {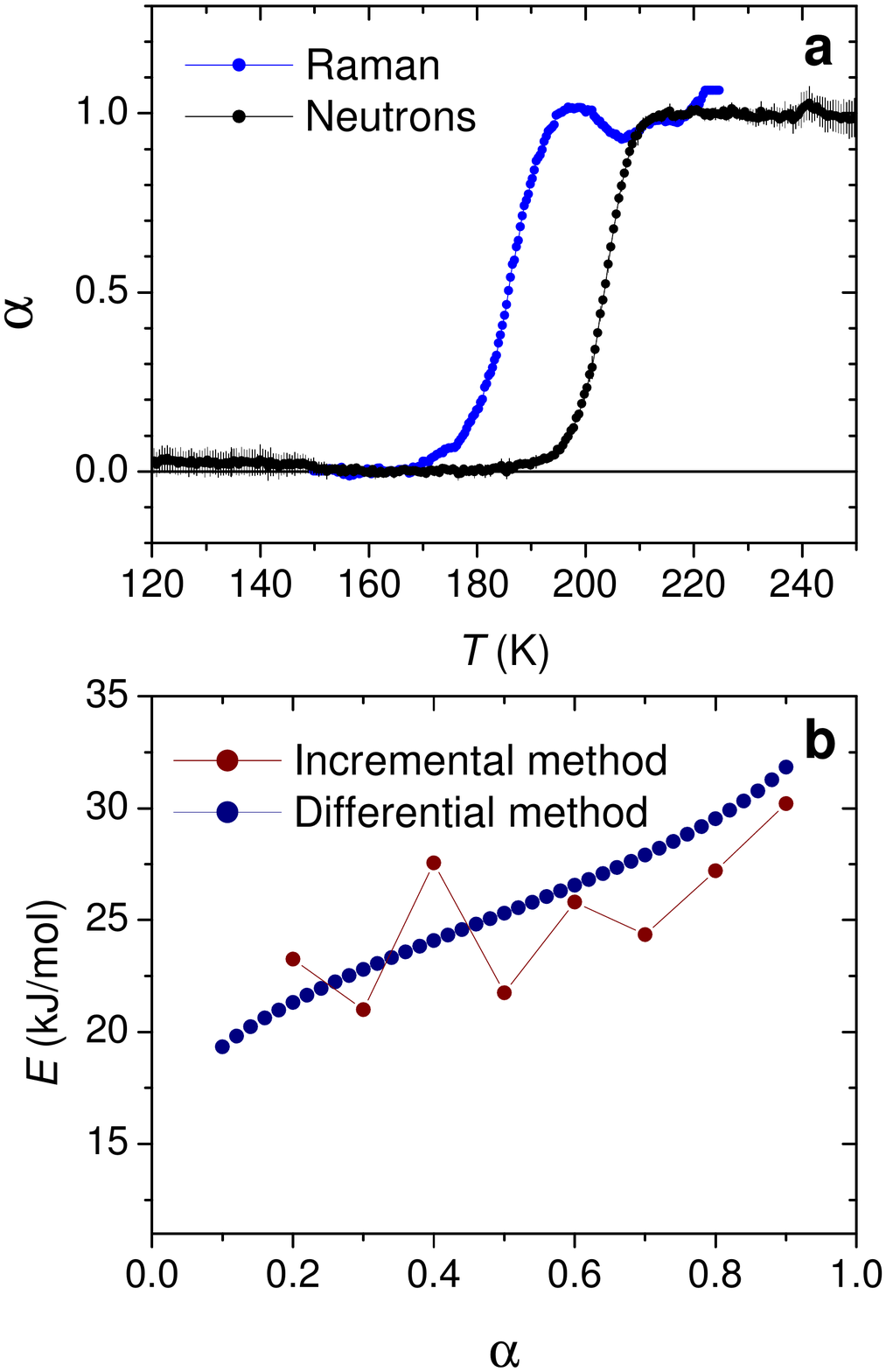}
 }
%

\end{center}
 \caption{(a) Evolution of the fraction of ice Ih, $\alpha$, measured during warming with rates respectively $\beta_1 = 0.0367$ K/min (blue, Raman spectroscopy data) and $\beta_2 = 0.1236$ K/min (black, neutron diffraction measurement\cite{del_Rosso20}).
 (b) Activation energy for the transformation, calculated with two different methods (see text).
 }
\label{f.ActivationEnergy}
\end{figure}

We have used the spectroscopic data described above also to study the kinetics of the Ic--Ih phase transformation in ice, during heating at the constant rate $\beta_{1} = 0.0367$ K/min (i.e. isochronal method\cite{Mittemeijer10}).
The intensity ratio $R(T)$ of the two peaks at frequencies $x_2$ and $x_1$ in the lattice band mirrors the phase transformation (see Fig. \ref{f.Lattice_XeR}(b)), and the fraction of ice Ih $\alpha(T)$, growing with temperature,  has been derived directly from it with a suitable normalization, 
\begin{equation}
\alpha (T) = 1-\frac{  R(T)-R_{lin}(T)  }    {  \langle R(T_0) \rangle -R_{lin}(T_0)  }
\end{equation}
where $R_{lin}(T)$ is the linear low-temperature extrapolation of $R(T)$ for ice Ih calculated from the data at $T>200$ K (gray straight line in Fig. \ref{f.Lattice_XeR}(b)), and $\langle R(T_0) \rangle$ is the average of $R(T)$ in a suitable range before the transition, around $T_0 =160$ K.
This quantity is plotted in the upper panel of Fig. \ref{f.ActivationEnergy}, together with the same quantity obtained from our neutron diffraction experiment \cite{del_Rosso20} during a similar isochronal temperature program, but with a different heating rate $\beta_{2} = 0.1236$ K/min.
As expected, the faster heating ramp of the neutron experiment moves forward the temperature range of transformation of about 20 K compared to the slower Raman experiment.

From a general point of view, such a phase transformation is a composite phenomenon, that involves at least three, generally overlapping, mechanisms, i.e. nucleation, growth and impingement\cite{Mittemeijer10, Guan18}. However, one can derive the effective parameters describing the transformation kinetics from non-isothermal data taken at different heating rates, even without adopting a specific kinetic model, if one relies on the so-called isoconversional method \cite{Simon04,Vyazovkin18}. 
In this framework, the conversion rate is assumed to depend only on temperature and converted fraction $\alpha$, according to 
\begin{equation}
\frac{\ud\alpha}{\ud t} =k(T)f(\alpha)=A \exp\left( -\frac{E}{RT} \right) f(\alpha)
\label{e.iso}
\end{equation}
where $f(\alpha)$ is called reaction model or conversion function.
Here we have expressed the rate constant $k(T)$ in the common Arrhenius form, being $E$ the activation energy and  $R$ the gas constant.
Since the experimentally accessible quantity is $\alpha(T)$, and not its time derivative, a direct application of Eq.~\ref{e.iso} requires either the computation of $\ud\alpha/\ud t$ from the experimental data (often flawed by numerical noise) or the integration of Eq.~\ref{e.iso}, (which requires an approximation of a non-analytical integral), leading, respectively, to the differential or integral method\cite{Simon04}.
Even if the determination of the activation energy of the process is affected by uncertainties when only few different isochronal processes are considered, we have pursued the derivation of $E$, applying both the differential\cite{Friedman64,Friedman69} and the advanced incremental integral method\cite {Vyazovkin01,Vyazovkin08}. In the first case, after the determination of the two sets of data $\left(\frac{\ud\alpha_{1}}{\ud t}\right)_{\alpha}$ and $\left(\frac{\ud\alpha_{2}}{\ud t}\right)_{\alpha }$ from the fit of the experimental Raman (1) and neutron (2) results, the activation energy $E=E_{\alpha}$, for different values of $\alpha$, has been readily obtained from the equation
\begin{equation}
E_{\alpha} = R \left( \frac{1} {T_2(\alpha)} - \frac{1}{T_1(\alpha)}  \right)^{-1}
 \ln \left[           \left( \frac{\ud\alpha_{1}}{\ud t} \right)_{\alpha} \left/
                       \left( \frac{\ud\alpha_{2}}{\ud t} \right)_{\alpha}   \right.    \right]
\label{e.differ}
\end{equation}
which derives directly from Eq.~\ref{e.iso}.
Here, $ T_1(\alpha)$ and $ T_2(\alpha)$ represents the temperatures at which the fraction $\alpha$ is attained in the Raman and neutron experiment, respectively.
Alternatively, within the advanced incremental method and considering $E_{\alpha}$ constant in the temperature interval $(T_i(\alpha - \Delta \alpha),T_i(\alpha))$ in both experiments ($i=1,2$), one obtains the activation energy as the values that minimizes the function $\Phi(E_{\alpha})$ :
\begin{equation}
\Phi(E_{\alpha}) = \frac{\beta_2 I(E_{\alpha}, T_{1}(\alpha))}{\beta_1 I(E_{\alpha}, T_{2}(\alpha))} + \frac{\beta_1 I(E_{\alpha}, T_{2}(\alpha))}{\beta_2 I(E_{\alpha}, T_{1}(\alpha))},
\end{equation}
where the integral 
\begin{equation}
I(E_{\alpha}, T_i(\alpha)) = \int_{T_{i}(\alpha - \Delta \alpha)}^{T_{i}(\alpha)} \exp \left(-\frac{E_{\alpha}}{R T} \right) \ud T
\end{equation}
has been resolved by a numerical method. 
The results for $E_{\alpha}$ obtained with the two methods, plotted and compared in Fig. \ref{f.ActivationEnergy}(b), are in a good agreement and show a slight increase in both cases. 
The dependence of the activation energy on $\alpha$ is a common outcome of the effective kinetics methods, since the different mechanisms occurring in the phase transformation are usually not well separated in the time-temperature domain. 
To our knowledge, the only previous determination of the activation energy, though relative to the transition ice Isd to ice Ih, is that reported in Ref. \onlinecite{Sugisaki68}, where a two step process was inferred from the experimental data, the second of which, occurring around 190 K, is characterized by an activation energy of about 45 kJ/mol. The two step process observed in Ref.  \onlinecite{Sugisaki68} is probably due to the presence of significant amount of stacking disorder in their starting sample of supposed cubic ice.
The fact that in our case the effective activation energy is almost constant let us think that this is a single step process, characterized by an averaged value of $E_{\alpha}$ equal to about 25 kJ/mol.
Other kinetic studies reported in literature did not reach conclusive determinations because of the stacking disorder present in the initial sample\cite{Hansen08b}.

In conclusion, this study has faced the characterization of the ice Ic--Ih transition from the new perspective of the vibrational dynamics. The continuous acquisition of the Raman spectra of the lattice and OH stretching vibrational modes performed during the transition has evidenced irreversible changes in the spectroscopic features of both bands, clearly ascribable to the crystalline symmetry change. The high sensitivity of the Raman apparatus has allowed to quantitatively measure the thermal behavior of the spectral characteristic parameters, identifying the overall shape of the lattice band and the width of the major peak in the OH stretching band as the main signatures to distinguish the two phases. Although we have provided an estimation for the activation energy of the Ic--Ih transformation, a more accurate characterization of the kinetics of such a thermally activated process should involve a series of both isothermal and isochronal anneals at different temperatures and heating rates, respectively. In this perspective, further kinetics studies of the Ic-Ih transformation are desirable. These, however, will involve the disruption of the sample after each long series of measurements, making this task quite challenging in terms of the experimental resources.

\begin{acknowledgement}

We are grateful to Mr. Andrea Donati (CNR-IFAC, Sesto Fiorentino, Firenze) for his fundamental technical support for setting up the high-pressure apparatus. L. Ulivi, M. Celli and L. del Rosso acknowledge also the support from the Fondazione Cassa di Risparmio di Firenze under the contract "Grandi Attrezzature 2019- HYDRO10000" (2019-0244).

\end{acknowledgement}

\begin{suppinfo}

The Supporting Information (SI) is available free of charge at https://pubs.acs.org/doi/.
\\
Detailed description of the synthesis procedure of the starting samples and insights in the fitting procedure used in the data analysis;
\\
Accurate measurements of the frequency position of the OH stretching vibration in ice I and comparison with the current literature.

\end{suppinfo}

\end{document}